\documentclass[11pt,a4paper]{emulateapj}

\usepackage{epsfig}
\usepackage{amsmath}
\usepackage{natbib}

\begin{document}

\title{Characterization of the Nearby L/T Binary Brown Dwarf 
WISE\,J104915.57$-$531906.1 at 2 Parsecs from the Sun\altaffilmark{1}}


\author{
A.\,Y.~Kniazev\altaffilmark{2,3},
P.~Vaisanen\altaffilmark{2,3},
K.~ Mu\v{z}i\'{c}\altaffilmark{4},
A.~Mehner\altaffilmark{4},
H.M.J.~Boffin\altaffilmark{4},
R.~Kurtev\altaffilmark{5},
C.~Melo\altaffilmark{4},
V.\,D.~Ivanov\altaffilmark{4},
J.~Girard\altaffilmark{4},
D.~Mawet\altaffilmark{4},
L.~Schmidtobreick\altaffilmark{4},
N.~Huelamo\altaffilmark{6},
J.~Borissova\altaffilmark{5},
D.~Minniti\altaffilmark{7},
K.~Ishibashi\altaffilmark{8},
S.\,B.~Potter\altaffilmark{2},
Y.~Beletsky\altaffilmark{9},
D.\,A.\,H.~Buckley\altaffilmark{3},
S.~Crawford\altaffilmark{2,3},
A.\,A.\,S.~Gulbis\altaffilmark{2,3},
P.~Kotze\altaffilmark{2,3},
B.~Miszalski\altaffilmark{2,3},
T.\,E.~Pickering\altaffilmark{2,3},
E.~Romero Colmenero\altaffilmark{2,3},
T.\,B.~Williams\altaffilmark{2,10,11}
}

\altaffiltext{1}
{Based on observations made with the Southern African Large
Telescope (SALT).}

\altaffiltext{2}{South African Astronomical Observatory, PO Box 9, 
7935 Observatory, Cape Town, South Africa}
\altaffiltext{3}{Southern African Large Telescope Foundation, PO 
Box 9, 7935 Observatory, Cape Town, South Africa}
\altaffiltext{4}{European Southern Observatory, Ave. Alonso de 
Cordova 3107, Casilla 19001, Santiago 19, Chile}
\altaffiltext{5}{Departamento de F\'isica y Astronom\'ia, 
Universidad de Valparaiso, Av. Gran Breta\~na 1111, Playa Ancha, 
5030, Casilla, Chile}
\altaffiltext{6}{CAB (INTA-CSIC), LAEFF, P.O. Box 78, E-28691 
Villanueva de la Ca\~nada, Madrid, Spain}
\altaffiltext{7}{Departamento Astronom\'ia y Astrof\'isica, 
Pontificia Universidad Cat\'olica de Chile, Av. Vicu\~na Mackenna 
4860, Santiago, Chile and Vatican Observatory, V00120, Vatican 
City State}
\altaffiltext{8}{Nagoya University, Japan}
\altaffiltext{9}{Las Campanas Observatory, Carnegie Institution 
of Washington, Colina el Pino, Casilla 601 La Serena, Chile}
\altaffiltext{10}{Department of Astronomy, University of Cape 
Town, Cape Town, South Africa}
\altaffiltext{11}{Department of Physics and Astronomy, Rutgers, 
the State University of New Jersey, Piscataway, NJ 08854, USA}

\begin{abstract}
WISE\,J104915.57$-$531906.1 is a L/T brown dwarf binary
located 2\,pc from the Sun. The pair contains the closest known 
brown dwarfs and is the third closest known system, stellar or 
sub-stellar. We report comprehensive follow-up observations of
this newly uncovered system. We have determined the spectral types
of both components  (L8$\pm$1, for the primary, agreeing with
the discovery paper; T1.5$\pm$2 for the secondary, which was lacking 
spectroscopic type determination in the discovery paper)
and, for the first time, their
radial velocities (V$_{rad}$$\sim$23.1, 19.5\,km\,s$^{-1}$) using 
optical spectra obtained at the Southern African Large Telescope 
(SALT) and other facilities located at the South African 
Astronomical Observatory (SAAO). The relative radial velocity of 
the two components is smaller than the range of orbital velocities 
for theoretically predicted masses, implying that they form a 
gravitationally bound system. We report resolved near-infrared 
$JHK_S$ photometry from the IRSF telescope at the SAAO which yields 
colors consistent with the spectroscopically derived spectral types. 
The available kinematic and photometric information excludes the 
possibility that the object belongs to any of the known nearby young 
moving groups or associations. Simultaneous optical polarimetry 
observations taken at the SAAO 1.9-m give a non-detection with an 
upper limit of 0.07\%. For the given spectral types and absolute 
magnitudes, 1\,Gyr theoretical models predict masses of 
0.04--0.05\,M$_{\odot}$for the primary, and 0.03--0.05\,M$_{\odot}$ 
for the secondary.
\end{abstract}


\section{Introduction}\label{sec:intro}

Nearby stars are easy to identify from their high proper motion (PM)  
which can reach many tenths of arcseconds per year. However, the
confusion against the crowded Milky Way background can make such 
objects hard to identify near the Galactic plane. The extreme red 
colors of late-type objects make their detection by the optical 
surveys challenging. The last few decades have seen great 
improvements with both these issues. A number of projects have
generated rich data sets for PM searches in the near-infrared (NIR)
and mid-infrared: Two Micron All-Sky Survey \citep[2MASS;][]{skr06}, 
the Deep Near-Infrared Survey of the Southern Sky 
\citep[DENIS;][]{ep99}, and the United Kingdom Infrared Telescope 
Infrared Deep Sky Survey \citep[UKIDSS;][]{law07}. Most recently, 
the Wide-field Infrared Survey Explorer \citep[{\it WISE};][]{wri10} 
imaged the entire sky at 3.4, 4.6, 12, and 22\,$\mu$m. This mission 
is particularly sensitive to sub-stellar objects, and it obtained 
observations over many epochs separated by 0.5-1\,yr. 

\citet{Luh13} used these multi-epoch WISE observations to search for 
objects with red colors and high-PM. He identified 
WISE\,J104915.57$-$531906.1 (hereafter W10$-$53) with 
$\mu$$\sim$3\arcsec\,yr$^{-1}$. Follow-up observations showed two
objects at the location of W10$-$53 and spectroscopy of the primary 
indicated an L8 spectral type. A parallax based on WISE and a number
of older surveys placed W10$-$53 at a distance of $\sim$2\,pc. This 
makes it the third closest system to the Sun, after 
Proxima/$\alpha$\,Cen and Barnard's star \citep{bar16}. These are 
now the closest brown dwarfs (BDs), usurping UGPS\,0722$-$05 \citep{Luc10} from 
this position. Additional archival detections were reported by
\citet{Mam13}. He used kinematic considerations to conclude that the
system probably belongs to the thin disk and is unlikely to be 
younger than 10$^8$\,yr. \citet{gandhi13} reported X-ray
non-detection setting an X-ray-to-bolometric-luminosity limit of 
W10$-$53 to log(L$_{\tt 0.2-3\,keV}$/L$_{\tt bol}$)$<$$-$4.6, 
consistent with this age.

W10$-$53, together with the other recently discovered nearby BDs
\citep{Luc10,Art10}, can constrain the local BD density and offers
us an opportunity to study these cool objects in detail, search for 
planets around them, and even resolve their surfaces with future 
interferometric instruments. 
We carried out optical spectroscopy, NIR imaging, and optical 
polarimetry of W10$-$53 with SALT and other facilities at the SAAO 
to obtain spectral types of both components, measure their radial 
velocities, test if the system may belong to a nearby moving group 
or association, and look for the presence 
of scattering dust.

\section{Observations and data reduction}\label{sec:obs_data_red}

\subsection{SALT Optical Spectroscopy}

\begin{figure}
\begin{center}
\includegraphics[angle=0,width=8.6cm,clip=]{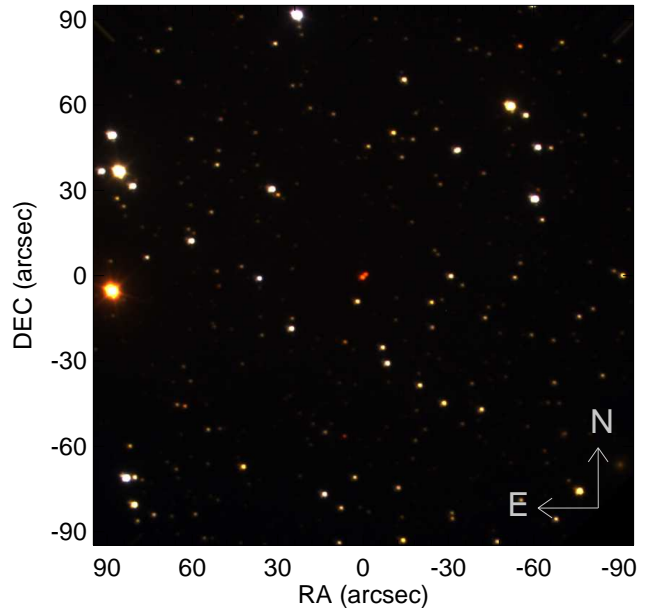}
\caption{Three-color optical image of the W10$-$53 field highlighting the 
extremely red color of the binary. Three 90\,sec exposures taken with the
RSS at SALT at  RA = 10h49m15.57s, DEC = $-$53\arcdeg19\arcmin06.1\arcsec\
were combined: red, green, and blue correspond to 100--200\,\AA\ wide 
filters centered at 8175, 7260, and 5060\AA, respectively.}
\label{fig:acqimage}
\end{center}
\end{figure}

Long-slit spectra of W10$-$53 were obtained with the Robert Stobie 
Spectrograph \citep[RSS;][]{Burgh03,Kobul03} at the Southern African 
Large Telecope \citep[SALT;][]{Buck06,Dono06} in Sutherland, South
Africa. The RSS uses a mosaic of three 2048$\times$4096 CCDs. The
spatial scale was 0.253\arcsec\,pix$^{-1}$, after binning by a factor 
of 2. The 0.6\arcsec\ wide slit was rotated to a
position angle of 133$\degr$ to observe both objects simultaneously. 
SALT makes use of an Atmospheric Dispersion Compensator, ensuring 
that there were no color dependent slit losses. The PG1800 grating 
was used on March 12, 2013 to measure the radial velocities because
it provides the highest spectral resolution at red wavelengths, 
resulting in spectral coverage of 7870--8940\,\AA\ and spectral 
resolution of 0.97\,\AA\ (0.33\,\AA\ per binned pixel). To improve 
the spectral typing, a wider range 6700--9670\,\AA\ was observed on 
March 16, 2013, with the PG900 grating, providing 2.19\,\AA\ 
(1.89\,\AA\ per binned pixel) resolution. A single 600\,sec spectrum 
was taken in each set up. The seeing during both observations was 
1.3--1.4\arcsec\ and the two components, separated by $\sim$1.5\arcsec, 
were resolved on the acquisition images (Fig.\,\ref{fig:acqimage}). A 
Neon lamp arc spectrum and a set of Quartz Tungsten Halogen (QTH) 
flats were taken immediately after the science frames. A 
spectrophotometric standard star, CD$-$32$\degr$9927, was observed 
for both setups.

The overscan, gain, cross-talk corrections, and mosaicing were done using 
the SALT data pipeline, PySALT \citep{Cr10}. The red end ($>$8400\,\AA) of 
the spectra suffer from significant fringing effects, but we were able to 
remove them using the QTH flats. MIDAS\footnote{Munich Image Data Analysis 
System is distributed by ESO.} and routines from the {\it twodspec} package 
in IRAF\footnote{IRAF is distributed by the NOAO, which is operated by the 
AURA under cooperative agreement with the NSF.} were used for wavelength 
calibration, frame rectification, and background subtraction of the 2D 
spectrum \citep{Kn08}. The derived internal error for the wavelength 
calibration is $\sigma$=0.03\,\AA\ throughout the wavelength range. This 
was verified against the numerous night sky lines in this wavelength range. 
Velocities were then corrected for heliocentric motion. Absolute flux 
calibration is not feasible with SALT because the unfilled entrance pupil 
of the telescope moves during the observation. However, a relative flux 
correction to recover the spectral shape was done using the observed 
spectrophotometric standard.

\begin{figure*}
\begin{center}
\includegraphics[angle=0,width=18.5cm]{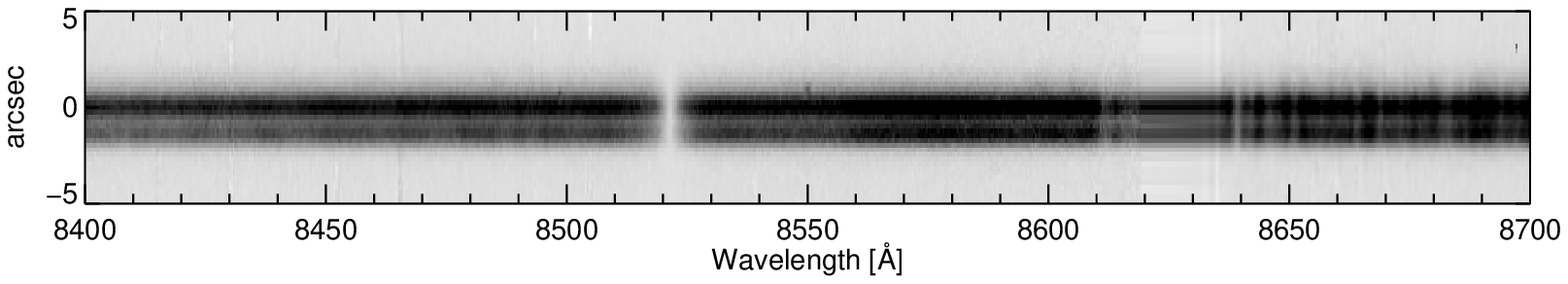}
\includegraphics[angle=-90,width=16.0cm,clip=]{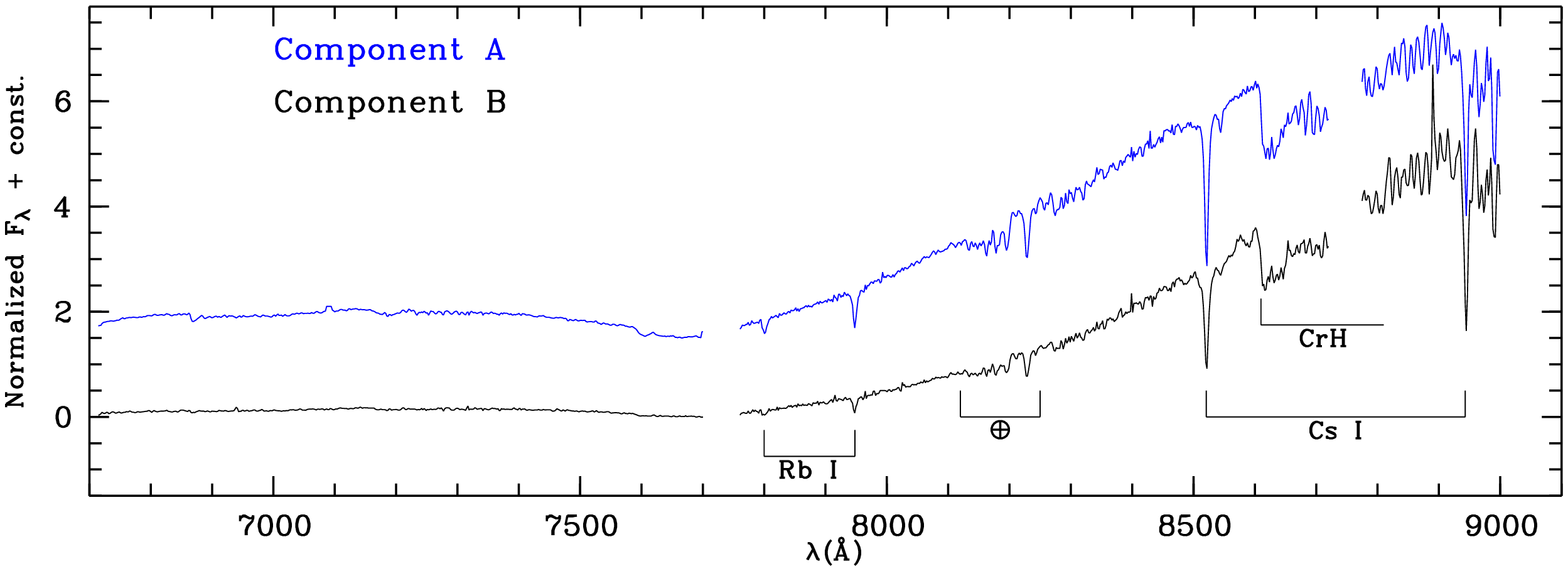}
\caption{
{\it Top:} Part of the wavelength range of the reduced 2D spectrum of 
W10$-$53 taken with the PG1800 grating, showing the spatially resolved 
components of the binary. Component A, brighter in the optical, is at 
the top. The Cs{\sc I} absorption line is clearly seen at 8521\,\AA, the 
CrH bandhead at 8610\,\AA\ and the gap between RSS detectors at 
8620--8635\,\AA\ are clearly seen.
{\it Bottom:} The extracted 1D spectra of both components obtained with 
the PG900 grating. The main spectral features are marked.
}
\label{fig:spec}
\end{center}
\end{figure*}

The top panel of Fig.\,\ref{fig:spec} shows a section of the fully reduced 
PG1800 2D spectrum demonstrating that we were successful in spatially 
separating the binary components. The 1D spectra were extracted with 5 
pixel ($\sim$1.3\arcsec) apertures. We optimized the width and the location 
of the apertures to minimize the cross-contamination between the two 
companions: the spectrum of B contains less than 6\% of the light from A, 
and the spectrum of A contains less than 3\% of the light from B. The lower 
panel of Fig.\,\ref{fig:spec} shows the wider wavelength range taken with 
the PG900 grating.

\subsection{IRSF NIR imaging}

W10$-$53 was imaged with the Simultaneous-Color InfraRed Imager for 
Unbiased Survey \citep[SIRIUS;][]{Nagayama03} on the Infrared Survey 
Facility (IRSF) 1.4-m telescope in Sutherland on March 16, 2013, under
clear conditions and $\sim 0.8$\arcsec\ $J$-band seeing.
SIRIUS has three 1024$\times$1024 HgCdTe detectors and it splits the 
incoming light by two dichroics for simultaneous $JHK_S$ observations. 
The scale is 0.45\arcsec\,px$^{-1}$, yielding 
$\sim$7.7$\times$7.7\,arcmin field of view. We took a set of 10 
dithered images with 5\,sec exposures for the individual frames. The 
data were reduced with the SIRIUS 
pipeline\footnote{The pipeline can be retrieved at
http://www.z.phys.nagoya-u.ac.jp/$\sim$nakajima/sirius/software/software.html.}.

The first reduction steps were flat-fielding and subtraction of dark current and sky background. 
Then, the ten dithered frames were aligned and combined into a final image. 
The astrometric calibration was derived from 2MASS stars in the field. 
Forty-nine of the 2MASS stars, selected to have no close neighbors, were used to 
determine the flux zero points. These stars span wide color and magnitude 
ranges (0.1$\leq$$J$$-$$K_{\rm S}$$\leq$1.5\,mag,
9.6$\leq$$K_{\rm S}$$\leq$15.6\,mag; but note that component A is 
somewhat outside the color range). 
The SIRIUS filter system is unique so we transferred the measured 
instrumental magnitudes into the 2MASS system. The color terms of our 
transformations are identical, within the uncertainties, to those of 
\citep{Kuchi08}. A least-squares fit to the transformations yielded 
an absolute photometric uncertainty below 3\%. 

The stellar photometry was carried out with {\sc ALLSTAR} in {\sc DAOPHOT 
II} \citep{stet87}. The PSF width of $\sim$1.0-1.1\arcsec\ and the 
$\sim$1.5\arcsec\ separation between the two components allowed us to 
achieve fitting errors of $\sim$0.008\,mag. The apparent magnitudes of 
the two components are reported in Table\,\ref{table1}. We list for 
comparison their combined magnitudes which are in excellent agreement 
with 2MASS and DENIS data.

\begin{table}
\centering
\caption{IRSF Photometry for W10$-$53 from March 16, 2013.}
\label{table1}
\begin{tabular}{cccc}
\hline
Band        & Component A$^a$  & Component B$^a$  &  Combined$^a$    \\
\hline
$J$         & 11.511$\pm$0.028 & 11.233$\pm$0.028 & 10.611$\pm$0.028 \\
$H$         & 10.396$\pm$0.026 & 10.369$\pm$0.028 &  9.634$\pm$0.026 \\
$K_S$       &  9.559$\pm$0.029 &  9.767$\pm$0.029 &  8.901$\pm$0.029 \\
\hline
\multicolumn{4}{l}{$^a$ Uncertainties represent the formal Poisson
errors.}
\end{tabular}
\end{table}

\subsection{SAAO 1.9-m Optical Polarimetry}

W10$-$53 was observed with the HI-speed Photo-POlarimeter 
\citep[HIPPO;][]{pot10} on the 1.9-m telescope of the South African 
Astronomical Observatory in simultaneous linear and circular polarimetry 
and photometry mode (all-Stokes) on March 19, 2013, under moonless 
photometric conditions. Measurements were performed with the Kron-Cousin 
{\it I} filter, for a duration of 40\,min. Background sky measurements 
were acquired immediately prior to the observations. Polarized standards
HD\,298383 and HD\,80558 were used to calculate the position angle offsets 
and efficiency factors. The two components of W10$-$53 were unresolved by 
the instrument. No polarization was detected with a firm upper limit of 
0.07\,\%.

\section{Analysis}
\label{sec:analysis}

\subsection{Spectral Classification}
\label{sec:sp_type}

\begin{figure*}
\begin{center}
\includegraphics[angle=0,width=18.0cm]{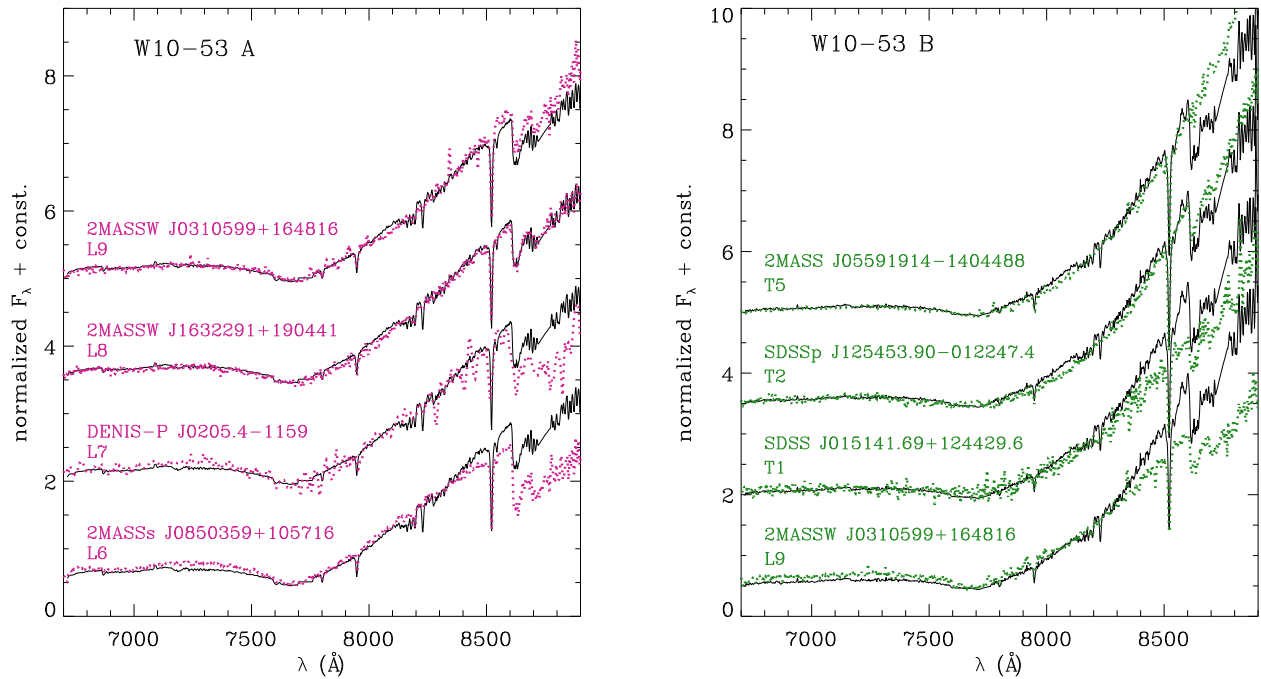}
\caption{Comparison of the W10$-$53\,A and B PG900 spectra (black) 
with various spectral templates (green and purple).
}
\label{fig:Spec_class}
\end{center}
\end{figure*}

To determine the spectral types of the two components, we created a 
spectral sequence of field L- and T- dwarf primary spectral standards 
from \citet{kirk99} and \citet{burgasser03}, adding extra L9
\citep{kirk00}, and T1 (classified in the NIR by \citealt{Mclean03})
templates\footnote{Data available at http://staff.gemini.edu/~sleggett/LTdata.html 
and http://www.iac.es/galeria/ege/catalogo\_espectral}.
Fig.\,\ref{fig:Spec_class} shows the PG900 spectra of W10$-$53\,A
and B (black) along with various spectral templates (purple for the 
primary and green for the secondary). Our spectra were smoothed by a 
boxcar function to a resolution of $\sim$3.3\,{\AA} and normalized at 
8200\,\AA.

The primary is best matched to L7--L9 spectra and we therefore classify
it as L8$\pm$1, same as \citet{Luh13}. Typing of the secondary is less 
straightforward, mainly because of the lack of reliable spectral templates 
of T-dwarfs in the optical. The spectrum appears to be bracketed by T1 and 
T2, and we assign T1.5$\pm$2 to this object. Our resolved NIR photometry 
yields colors consistent with the colors of BDs with the same spectral 
types from the Dwarf 
Archives\footnote{http://spider.ipac.caltech.edu/staff/davy/ARCHIVE/index.shtml}.
\citet{burgasser13} presented resolved NIR spectroscopy of W10$-$53,
deriving L7.5$\pm$0.9 and T0.5$\pm$0.7, for the primary and secondary, 
respectively, consistent with our spectral types.


The measured apparent magnitudes combined with the astrometrically 
calibrated absolute magnitudes of \citet{dup12} yield distances for 
the primary and the secondary of 2.5$\pm$0.3 and 2.0$\pm$0.2\,pc, 
respectively. Their consistency with the parallax of \citet{Luh13},
together with the typical magnitudes of the components, make it 
unlikely to find that W10$-$53 is a nearly-equal mass hierarchical 
BD system. However, the magnitudes must be treated with caution 
until further studies, because the photometric monitoring in I+$z$ 
by \citet{gillon13} yielded a quasi-periodic (P=4.87$\pm$0.01\,hr) 
variation of the secondary, probably produced by fast rotation and 
fast-changing cloud structure.

\subsection{Radial Velocity}\label{sec:Vrad}

We estimated the radial velocities for both components from the SALT 
medium-resolution spectra. The two most prominent absorption lines were 
used, identified in both spectra as Rb\,I at 7947.60~\AA\ and Cs\,I at 
8521.13~\AA\ \citep[Fig.\,\ref{fig:spec} and][]{kirk99}. To exclude
systematic shifts originating from the known RSS flexure, we calculated 
the line-of-sight velocity distributions along the slit using the 
method and programs described in \citet{Zasov00} where the nearest 
night-sky lines were used as references. The IRAF task {\tt splot} was 
used to calculate the centers of the absorption lines providing the 
measured radial velocities V$_{rad}$=23.1$\pm$1.1 and 
19.5$\pm$1.2\,km\,s$^{-1}$ for components A and B, respectively. These 
values incorporate a heliocentric correction of 7.4\,km\,s$^{-1}$, 
corresponding to the observing time of March 13, 2013, 01:14 UT. 
We cross-correlated the SALT medium-resolution spectra of component A
with component B to measure directly their relative velocity: 
2.5$\pm$1.9\,km\,s$^{-1}$, consistent with the difference between the
velocities given above: 3.6$\pm$1.6\,km\,s$^{-1}$.

As a double-check, we cross-correlated the spectra of the two 
components with a 1400\,K synthetic spectrum computed with the Phoenix 
simulator\footnote{http://phoenix.ens-lyon.fr/simulator/index.faces},
using the BT-Settl models \citep{allard03}. The cross-correlation,
performed with the IRAF task {\tt xcsao}, yielded 
22.0$\pm$4.0\,km\,s$^{-1}$ and 18.2$\pm$4.2\,km\,s$^{-1}$, in 
excellent agreement with the values obtained above. These estimates
are less reliable because the template refers to averaged parameters 
of both components, and they use all lines, not just the strongest 
ones, subjecting the cross-correlation to more noise.

The observed PM of the star translates to a transverse velocity of
28.4\,km\,s$^{-1}$, and the total relative velocity of the star with 
respect to the Sun is $\sim$36\,km\,s$^{-1}$.

\subsection{Polarization}\label{sec:Polar}

Linear polarization in cool sub-stellar objects is thought to arise in
disks, or due to dust in their atmospheres, combined with asymmetries 
caused by oblateness due to fast rotation or partial cloud coverage. The 
theoretical models predict that these mechanisms can cause polarization 
of up to 0.1\% in the optical \citep{Mar11}. Indeed, linear polarization 
reaching 0.2--1.7\,\% in $RI$-bands was found in some early-L BDs 
\citep{Tat09}. Our upper limit of 0.07\,\% argues against the presence 
of any of these polarizing mechanisms in W10$-$53.

\section{Summary and Conclusions}\label{sec:discussion}

We assign spectral type L8$\pm$1 to the primary and T1.5$\pm$2 to the
secondary. This classification is consistent with the NIR magnitudes 
of the two components, excluding the possibility that W10$-$53 is a 
nearly-equal mass, hierarchical BD system.

The relations between spectral type and effective temperature T$_{eff}$ 
of \citet{stephens09}, yield T$_{eff}$$\approx$
1350\,($\pm$60\,random,\,$\pm$100\,systematic)\,K for the primary, and 
1210\,($\pm$40\,random,\,$\pm$100\,systematic)\,K 
for the secondary. The random errors reflect the template matching, and 
the systematic ones reflect the calibration's rms. Compared with the
DUSTY \citep{chabrier00, baraffe02} and BT-Settl \citep{allard11} 
theoretical 1\,Gyr isochrones, these T$_{eff}$ values correspond to 
masses of 0.04--0.05\,M$_{\odot}$ for the primary, and 
0.03--0.05\,M$_{\odot}$ for the secondary. At 10\,Gyr their masses would 
still be in the sub-stellar regime: 0.065--0.072\,M$_{\odot}$ and 
0.06--0.07\,M$_{\odot}$, respectively. An evaluation of masses based on 
the NIR magnitudes alone yields similar ranges, albeit wider due to the 
larger uncertainties. With these masses and assuming a circular orbit 
with a radius equal to the projected separation at the parallax-derived
distance, we obtain a period in the range of 14-20\,years, and an orbital 
velocity of 4.6--6.5\,km\,s$^{-1}$. Our relative radial velocity between
the two components is 2.5--3.6\,km\,s$^{-1}$, implying that they are
indeed bound. The actual separation between them is possibly larger than 
the projected one or the orbit is very eccentric.

Combining the measured radial velocities (V$_{rad}$=23.1
$\pm$1.1,\,19.5$\pm$1.2\,km\,s$^{-1}$) with the distance and the PM from 
\citet{Luh13}, following \citet{JoSo87}, we obtain Galactic velocities: 
$U$=$-$17.8, $V$= $-$29.7, $W$=$-$6.5\,km\,s$^{-1}$. These are not 
compatible with any known nearby moving group or stellar association, 
furthering the initial assessment of \citet{Mam13}. In particular, the 
radial velocities we derive are too large by a factor of 3 to reconcile 
W10$-$53 with the 40 Myr-old Argus group. This is not surprising, as 
\citet{Mam13} already pointed out that the NIR photometry is not 
compatible with such a young age. The available measurements also 
exclude membership into the 30 Myr-old Car association \citep{Torres08}, 
unless the distance to W10$-$53 was shortened by more than 25\% (highly 
unlikely). A search through more than half a dozen Galactic
velocity catalogs yielded a single close match -- the M0 dwarf G\,171$-$22 
\citep{Wooley70}.\footnote{\citet{Egg62} lists $U$=$-$19.6, $V$=$-$29.3, 
$W$=$-$6.7\,km\,s$^{-1}$ for HD\,55383 but our re-calculation shows that 
the sign of $U$ is flipped.}
Many of the most recent PM and radial velocity catalogs don't include 
explicitly $UVW$, so our search is not conclusive, but it indicates that
W10$-$53 most likely lacks nearby bright co-moving companions, and its 
age remains poorly constrained.

\acknowledgements
Some observations reported in this paper were obtained with the Southern 
African Large Telescope (SALT). All SAAO and SALT co-authors acknowledge 
the support from the National Research Foundation (NRF) of South Africa.
RK and JB acknowledge partial support from FONDECYT through grants
No 1130140 and 1120601 respectively.
This research has benefited from the MLTY dwarf compendium at
DwarfArchives.org. We thank Adam Burgasser for making some template
spectra available to us, and the anonymous referee for the suggestions
that helped to greatly improve the paper.


\end{document}